# Key classification attack on block ciphers

Maghsood Parviz
Mathematical Sciences Department
Sharif University of Technology
Tehran, IRAN
maghsoudparviz@alum.sharif.ir

Saeed Mirahmadi
Pooyesh Educational Institute,
Qom, IRAN

Seyed Hassan Mousavi
Mathematical Sciences Department
Isfahan University of Technology
Isfahan, IRAN
shnmousavi_iut@yahoo.fr

*Abstract-* **In this paper, security analysis of block ciphers with key length greater than block length is proposed. When key length is significantly greater than block length and the statistical distribution of cipher system is like a uniform distribution, there are more than one key which map fixed input to fixed output. If a block cipher designed sufficiently random, it is expected that the key space can be classified into same classes. Using such classes of keys, our proposed algorithm would be able to recover the key of block cipher with complexity** $O(max\{2^n, 2^{k-n}\})$ **where n is block length and k is key length. We applied our algorithm to 2-round KASUMI block cipher as sample block cipher by using weakness of functions that used in KASUMI.**

## I. Introduction

In design of secure block cipher, one of the main steps is selection of key and input block length. There are some standard block ciphers like KASUMI, IDEA, 3DES, AES-256 which have block length smaller than key length. Since these ciphers have considered as pseudo-random generator, their key space can be classified into equivalency class of keys with the same number of elements. For instance, assuming randomness of KASUMI's output, the key space would have $2^{64}$ key classes which each contains $2^{64}$ different keys.

In our proposed attack algorithm, we assumed that there is an efficient algorithm which computes the equivalency class of an arbitrary key.

As an application of this method, we propose an efficient algorithm for computing equivalency class of arbitrary key for one-round and 2-round KASUMI block cipher which can be extended to full round KASUMI. Hence, in the rest, we give a short introduction to KASUMI block cipher.

KASUMI is modified version of the block cipher MISTY1 [1], which is optimized for hardware performance. In the past few years, it has received a lot of attentions from the cryptographic researcher. Kuhn introduced the impossible differential attack on 6-round KASUMI [2], which has been recently extended to 7-round KASUMI[3]. In 2007, a higher order differential attack has been published on 5-round KASUMI with practical complexity [4]. Since the key schedule of KASUMI is linear, many related-key attacks have published. A related-key differential attack on 6-round KASUMI has presented in 2002[5]. The first related-key attack on the full 8-round KASUMI was proposed by Biham et al.with $2^{76.1}$ encryptions [6], which was improved to a practical related-key attack on the full KASUMI by Dunkelman et al.[7].

However, assumption for these kinds of attacks is controlling over the differences of two or more related keys in all the 128 key bits. With this assumption, resulting attack is inapplicable in most real-world usage scenarios [8].

In this paper, we proposed a new attack based on relation between key length and block length of a block cipher. When key length is significantly greater than block length and cipher system has a statistical properties like an uniform distribution, there is a key classes with $2^{k-n}$ members K in which $E(P_0, K) = C_0$, for every fixed $(P_0, C_0)$ pair, where $E : \{0,1\}^n \times \{0,1\}^k \to \{0,1\}^n$ is corresponding encryption algorithm[9].

Using this observation, we construct an equivalency class of those keys satisfied in $E(P_0, K) = C_0$. In the rest of this paper, we assume that for every block cipher algorithm E, there is an algorithm which computes efficiently the equivalency class of an arbitrary key. Our method can be applied to block cipher like KASUMI, IDEA, AES-256, 3DES, etc.

In this paper, first we introduce our new method and then it will apply to reduced-round KASUMI block cipher. In section II, we introduce details of our method. In section III, the method has applied to KASUMI. In section V, we propose an algorithm for generating equivalency classes of KASUMI key space. Finally, after experimental results, we will give some suggestions and recommendations for future works.

## II. OUR PROPOSED ATTACK

Consider a block cipher algorithm $E : \{0,1\}^n \times \{0,1\}^k \to \{0,1\}^n$ in which n is block length and k is key length.

When key length is significantly greater than block length and the statistical distribution of cipher system is like a uniform distribution, there are $2^{k-n}$ keys K such that

$$E(P_0, K) = C_0$$

For every fixed $(P_0, C_0) \in \{0,1\}^n \times \{0,1\}^n$.

It simply can prove that the following relation is an equivalency relation.

$$K \sim K' \text{ iff } E(P_0, K) = E(P_0, K') = C_0$$

For all $(P_0, C_0) \in \{0,1\}^n$.

In the rest of the paper, $[K]_{P_0, C_0}$ is Equivalency relation for $K \in \{0,1\}^k$. For every block cipher algorithm E, suppose that there is an algorithm which efficiently computes the equivalency class of an arbitrary key.

Following algorithm will be able to recover the unknown key $K_0$ such that $E(P_0, K_0) = C_0$, for known $(P_0, C_0)$ with complexity $2^n$.

Algorithm 1:
1. Select the key K such that $E(P_0, K) = C_0$.

We use pseudo-randomness properties of encryption algorithm and it is expected that there is exactly one key such that $E(P_0, K) = C_0$ for every interval $[i \times 2^n, (i+1) \times 2^n]$ and $0 \le i \le 2^{k-n}$. For finding equivalence key, it would be sufficient to search whole keys in $[0, 2^n - 1]$. It can be done using rainbow-tables and time-memory-data-trade-off methods with complexity lower than $2^n$.

2. Since $K_0 \in [K]_{P_0, C_0}$, generate the keys of class $[K]_{P_0, C_0}$. If an arbitrary key decrypts at least three pair (cipher-text, plain-text) correctly, select it as main unknown key.

## III. A KEY CLASSIFICATION ATTACK ON KASUMI 2-ROUNDS BLOCK CIPHER

In this section, we apply our proposed method to 2-rounds KASUMI block cipher. By finding some weakness in FO and FL functions, we will improve our method by using them. In the rest, first we introduce briefly the general structure of KASUMI block cipher and then properties of functions that used in KASUMI are discussed.

### A. KASUMI block cipher

KASUMI is a block cipher used in the security of 3GPP systems such as UMTS, GSM and GPRS. Both the confidentiality (f8) and integrity function (f9) in UMTS are based on KASUMI. In GSM, KASUMI is used in the A5/3 algorithm for generating key stream and in GPRS in the GEA3 key stream generator.

KASUMI is an 8-round Feistel block cipher algorithm; it operates on 64 bit input to produce 64 bit output under a 128-bit key. In each round, there are two functions: the FO function which is a 3-round, 32-bit Feistel structure, and the FL function which receives 32-bit as input and produces 32-bit output. The order of using two mentioned functions in the cipher is affected by the round number. In the odd round the first function is FL and in the even round the FO function is used first.

The FO function also uses four-round Feistel FI function in a recursive structure. The FI function receives 16-bit as input and produce 16-bit as output. It uses two S-boxes S7 ( 7-bit to 7-bit permutation) and S9 (9-bit to 9-bit permutation).

Using a simple key schedule of figure 1, the round function uses a round key which consists of eight 16-bit sub keys derived from the original 128-bit key. The FL function uses 32-bit sub-keys $KL_{i,j}$ in round I where j=1 or 2. The FO function uses 96-bit sub-keys $KO_{i,j}$ and $KI_{i,j}$.

The 128-bit key $K$ is divided into eight 16-bit sub keys $K_i$:

$K = (k_1, k_2, k_3, k_4, k_5, k_6, k_7, k_8) \quad k'_i = k_i \oplus c_i$

| Round | $KL_{i,1}$ | $KL_{i,2}$ | $KO_{i,1}$ | $KO_{i,2}$ | $KO_{i,3}$ | $KI_{i,1}$ | $KI_{i,2}$ | $KI_{i,3}$ |
|---|---|---|---|---|---|---|---|---|
| 1 | $k_1 \lll 1$ | $k'_3$ | $k_2 \lll 5$ | $k_6 \lll 8$ | $k_7 \lll 13$ | $k'_5$ | $k'_4$ | $k'_8$ |
| 2 | $k_2 \lll 1$ | $k'_4$ | $k_3 \lll 5$ | $k_7 \lll 8$ | $k_8 \lll 13$ | $k'_6$ | $k'_5$ | $k'_1$ |
| 3 | $k_3 \lll 1$ | $k'_5$ | $k_4 \lll 5$ | $k_8 \lll 8$ | $k_1 \lll 13$ | $k'_7$ | $k'_6$ | $k'_2$ |
| 4 | $k_4 \lll 1$ | $k'_6$ | $k_5 \lll 5$ | $k_1 \lll 8$ | $k_2 \lll 13$ | $k'_8$ | $k'_7$ | $k'_3$ |
| 5 | $k_5 \lll 1$ | $k'_7$ | $k_6 \lll 5$ | $k_2 \lll 8$ | $k_3 \lll 13$ | $k'_1$ | $k'_8$ | $k'_4$ |
| 6 | $k_6 \lll 1$ | $k'_8$ | $k_7 \lll 5$ | $k_3 \lll 8$ | $k_4 \lll 13$ | $k'_2$ | $k'_1$ | $k'_5$ |
| 7 | $k_7 \lll 1$ | $k'_1$ | $k_8 \lll 5$ | $k_4 \lll 8$ | $k_5 \lll 13$ | $k'_3$ | $k'_2$ | $k'_6$ |
| 8 | $k_8 \lll 1$ | $k'_2$ | $k_1 \lll 5$ | $k_5 \lll 8$ | $k_6 \lll 13$ | $k'_4$ | $k'_3$ | $k'_7$ |

$x \lll i$ : $x$ rotate left by $i$ bites

Figure 1. Key schedule for KASUMI block cipher

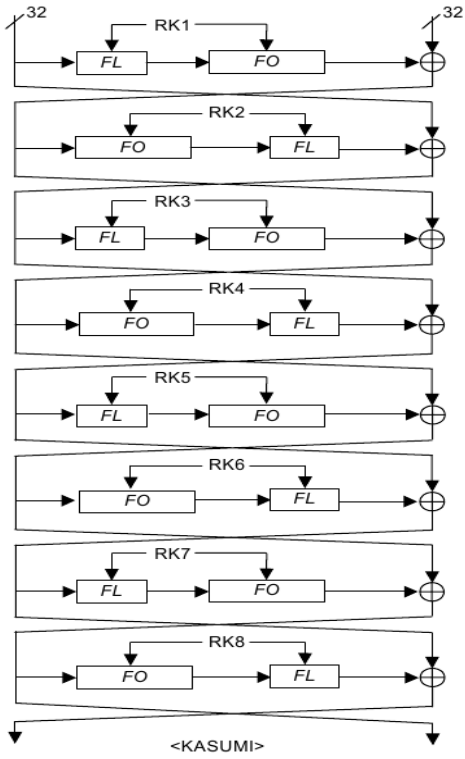

Figure 2. structure of KASUMI block cipher

We start with FI function which is the core function for giving non-linearity to the cipher algorithm.

B. *FI function properties*

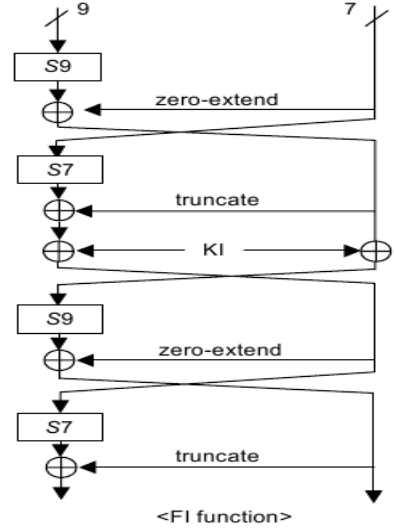

Figure 3. FI function

One of the main functions used in KASUMI block cipher is

$$FI : \{0,1\}^{16} \times \{0,1\}^{16} \to \{0,1\}^{16}$$

As figure 3 show, there are two -boxes S7 (7-bit to 7-bit permutation) and S9 (9-bit to 9-bit permutation). In the following, we give some properties for the FI function.

1. $FI(x, KI) = FI_2(KI \oplus FI_1(x))$

where $FI_1$ is above half of the FI function before the key KI affects and $FI_2$ is below half of the FI function after the key KI affects.

It is clear that $FI_1$ and $FI_2$ are invertible functions that are not depended on key.

2. Suppose that

$$FI(x, KI) = y \quad (1)$$

Where x, y and KI are input, output and key. Having two members of the set {x,y,KI}, the third one is uniquely computed efficiently. In fact, if {x,KI} or {y,KI} is known, using encryption and decryption algorithm, (1) holds.

If {x,y} is known, then

$$KI = FI_1(x) \oplus FI_2^{-1}(y)$$

3. $|\{(x, KI); FI(x, KI) = y\}| = 2^{32}$

for every $y \in \{0,1\}^{16}$.

4. Find KI and KI' for every $x, x' \in \{0,1\}^{16}$ such that

$$FI(x, KI) = FI(x', KI')$$

In fact, it is sufficient to find keys such that

$$KI \oplus KI' = FI_1(x) \oplus FI_1(x') \qquad (2)$$

It is clear that the number of keys which (2) hold are $2^{16}$.

5. There are some key pairs KI, KI' such that

$$FI(x, KI) = FI(x', KI') = y$$

For all $x, x', y \in \{0,1\}^{16}$.

Using the results of part 2, there are two keys pair with above properties.

## C. FO function properties

$FO: \{0,1\}^{32} \times \{0,1\}^{96} \to \{0,1\}^{32}$ is one of the core functions that used in KASUMI block cipher

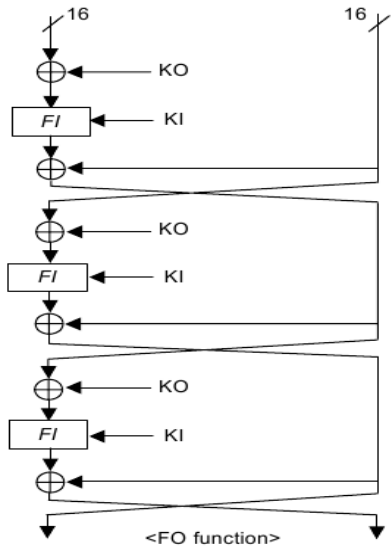

Figure 4.  FO function

Suppose $X = (X_L, X_R)$ and $Y = (Y_L, Y_R)$

are input and output of FO function, then

$$Y_L = X_R \oplus FI(X_L \oplus KO_1, KI_1)$$

$$\oplus FI(X_R \oplus KO_2, KI_2) \qquad (3)$$

$$Y_R = Y_L \oplus$$
$$FI(X_R \oplus FI(X_L \oplus KO_1, KI_1) \oplus KO_3, KI_3) \qquad (4)$$

Equation (4) can be rewritten as follow.

$$FI^{-1}(Y_L \oplus Y_R, KI_3) \oplus KO_3 =$$
$$X_R \oplus FI(X_L \oplus KO_1, KI_1) \qquad (5)$$

Finally, we have

$$X_R \oplus FI(X_L \oplus KO_1, KI_1) =$$
$$Y_L \oplus FI(X_R \oplus KO_2, KI_2)$$
$$= FI^{-1}(Y_L \oplus Y_R, KI_3) \oplus KO_3 \qquad (6)$$

When input and output of the FO function (X, Y) are known, then there are $2^{64}$ keys K such that FO(X,K)=Y.

For arbitrary values of $KI_1, KO_1, KI_2, KI_3 \in \{0,1\}^{16}$, $KO_2, KO_3$ can be computed efficiently and uniquely.

## D. FL function properties

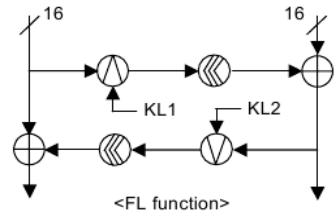

Figure 5.  FL function

FL function $FL: \{0,1\}^{32} \times \{0,1\}^{32} \to \{0,1\}^{32}$ is one the simple components in the KASUMI block cipher. We introduce some properties of FL function.

1. By fixing FL sub-keys, the function is one-to-one related to the input. But this isn't correct for the sub-keys by fixing the input.
2. If $X = (X_L, X_R)$, $Y = (Y_L, Y_R)$ are input, output of the FL function, and $KL = (KL_1, KL_2)$ is the corresponding sub-key, then

$$(Y_L \oplus X_L)^{>>1} = Y_R \vee KL_2 \quad (7)$$

$$(Y_R \oplus X_R)^{>>1} = X_L \wedge KL_1$$

Using (7), it is possible to classify all the FL sub-keys.

## IV. AN ALGORITHM FOR GENERATING EQUIVALENCE KEYS IN THE FIRST ROUND

In this section, using mentioned properties for function that used in KASUMI, we propose an efficient algorithm for finding class of keys which map fixed input $P_0$ to fixed output $C_0$ for $P_0, C_0 \in \{0,1\}^{64}$ ($C_0$ is the output of the first round before substitution of two halves)

Algorithm 2:

Given $P_0 = (P_L, P_R), C_0 = (C_L, C_R)$

Find every $K \in \{0,1\}^{128}$ such that

$KASUMI_{1-round}(P_0, K) = C_0$

For $(KL_1, KL_2, KO_1, KI_1, KI_2, KI_3) \in \{0,1\}^{96}$ do

find $(KO_2, KO_3) \in \{0,1\}^{32}$ such

that $P_R \oplus FO(FL(P_L, KL), KI, KO) = C_R$

Using FO function properties, we will be able to search efficiently desired key values.

$$KO = (KO_1, KO_2, KO_3), \quad KI = (KI_1, KI_2, KI_3).$$

There are $2^{96}$ different keys K that $KASUMI_{1-round}(P_0, K) = C_0$ for fixed $P_0, C_0 \in \{0,1\}^{64}$. The sub-keys values $(KO_2, KO_3) \in \{0,1\}^{32}$ can be computed efficiently for every guess of sub-keys $(KL_1, KL_2, KO_1, KI_1, KI_2, KI_3) \in \{0,1\}^{96}$. Therefore current algorithm is efficient since the complexity of algorithm is $2^{96}$ and this is the best complexity with regard to the number of key bits.

## V. AN ALGORITHM FOR GENERATING EQUIVALENCE KEYS IN THE SECOND ROUND

We extend our algorithm to 2-rounds KASUMI. It is expected that there are $2^{64}$ members in $[K]_{P_0, C_0}$ for fixed $P_0, C_0 \in \{0,1\}^{64}$ ($C_0$ is the output of the second round before substitution of two halves). Since we are in the initial rounds of KASUMI and the statistical properties of the cipher algorithm are not complete, the number of elements of class $[K]_{P_0, c_0}$ may not be similar to that we expected.

According to cipher structure, there are some relations between input and output of 2-round KASUMI block cipher as follow.

$$C_L \oplus P_R = FO(FL(P_L, RKL_1), RKO_1, RKI_1)$$

$$C_R \oplus P_L = FL(FO(C_L, RKO_2, RKI_2), RKL_2)$$

Where $RKL_1$ is the sub-key of first round for KL. Consequently, it needs to solve a system of equations for recovering unknown sub-keys.

Based on key schedule, if $K_1$ is known, the sub-key $KL_1$ of the first round and $KI_3$ of the second round, etc will be known, where $K = (K_1, K_2, K_3, K_4, K_5, K_6, K_7, K_8)$ is the main key of the cipher for producing key schedule.

In our algorithm, we guess FL function sub-keys for the first two round, then using FO function equations (3-6), the other sub-keys can be computed. In fact, in the first two rounds, these FL sub-keys are $K_1, K_2, K_3, K_4$. We represent our proposed algorithm in which the unknown key values show by capital letters.

Algorithm 3:

Given $P_0 = (P_L, P_R), C_0 = (C_L, C_R)$

Find every $K \in \{0,1\}^{128}$ s.t

$KASUMI_{2-round}(P_0, K) = C_0$

For $(k_1, k_2, k_3, k_4) \in \{0,1\}^{64}$ do

$a := FL(P_L, RKL_1);$

$b := C_L \oplus P_R;$

$c := C_L;$

$d := FL^{-1}(C_R \oplus P_L, RKL_2);$

Solve the following equations,

$$a_R \oplus FI(a_L \oplus k_2^{<<5}, K'_5) = b_L \oplus FI(a_R \oplus K_6^{<<8}, k'_4)$$
$$= K_7^{<<13} \oplus FI^{-1}(b_L \oplus b_R, K'_8)$$

$$c_R \oplus FI(c_L \oplus k_3^{<<5}, K'_6) = d_L \oplus FI(c_R \oplus K_7^{<<8}, K'_5)$$
$$= K_8^{<<13} \oplus FI^{-1}(d_L \oplus d_R, k'_1)$$

It can be shown that if we would be able to find some part of keys, the other part of key can be computed uniquely. In this algorithm, when we find $K_6$, sub-key $K_5$ can be computed from the first equation of algorithm 2. Using the second equation of algorithm 2, $K_7, K_8$ can be computed.

For correctness, these sub-key values should be check with the additional check equation in the second part of the first equation of algorithm 2.

It should be noted that, in algorithm 2, solving strategy can be every appropriate strategy for solving equations. For example we can replace this strategy with exhaustive search or algebraic methods to find $K_6$ and then compute the rest of sub-key values.

Complexity of the algorithm 2 depends on solving strategy. In the worst case, if exhaustive search method is used, the complexity of algorithm will be $2^{80}$ which is the best one with regards to number of key bits.

## VI. EXPERIMENTAL RESULTS

Since KASUMI block cipher has 64-bit block length and 128-bit key length, based on birthday paradox, when a fixed input encrypts with $2^{32}$ different keys, there will be at least two equal cipher-text with probability more than ½. With P=0 (plain-text with 64 bit zero), there are at least two keys that encryption of P lead to same cipher-text.

Using such keys, we get output of the sixth round KASUMI; we wrote equations for last two round and the results are compared. We run our algorithm on 4,6,7-rounds KASUMI separately and obtain some equivalence keys. For 6-round KASUMI, we found three keys with same cipher-text. It can be inferred that 6-round KASUMI shows more non-random properties and it can be vulnerability for KASUMI block cipher, since we expected two keys with same cipher with considering $2^{32}$ different keys.

## VII. CONCLUSIONS

In this respect, we proposed a new attack on block cipher algorithms with key length greater than block length, if it would be possible to efficiently classify the key space of corresponding block cipher. In fact, the complexity of attack will be $O(max\{2^n, 2^{k-n}\})$ where n is block length and k is key length.

Some recommendations for continuing this work are as follow.

1. Extend our algorithm to higher rounds of KASUMI block cipher.
2. Solving algebraic equations obtained by 2-rounds KASUMI in an efficient way.
3. Compute $p(C_i = C'_i | C_j = C'_j)$ for $i > j$ where $C_i, C_j$ are the output of i'th-round KASUMI with same plain-text and different keys.
4. Cryptanalysis of 4-rounds KASUMI block cipher with complexity of $2^{64}$.
5. Classification of the key space of other cipher algorithms such as IDEA, 3DES, AES-256 with key length greater than block length.
6. Combining this method with other known attack to achieve a near practical attack algorithm.

APPENDIX 1.

Plain-text=0
Key=0xF1D941159CA8B6238135DACB8A370940
Cipher-text=0x2DBCDA8D84CDAD86

--->> c1: left=0, right=db16eed5

--->> c2: left=db16eed5, right=48d17eb6

--->> c3: left=48d17eb6, right=2ebddad4

--->> c4: left=2ebddad4, right=7b006cf8

--->> c5: left=7b006cf8, right=d8805ffd
--->> c6: left=d8805ffd, right=9f570e58
--->> c7: left=9f570e58, right=84cdad86
--->> c8: left=84cdad86, right=2dbcda8d

Where "ci" is the output of round I in the KASUMI block cipher.

Plain-text=0
Key= 0xCAFF6AC383136437A70C4560AC98CE9F

Cipher-text= 0x2DBCDA8D84CDAD86

--->> c1: left=0, right=aa108129
--->> c2: left=aa108129, right=ec2e85a9
--->> c3: left=ec2e85a9, right=309e5e7b
--->> c4: left=309e5e7b, right=8f1313fb
--->> c5: left=8f1313fb, right=2b23dcc6
--->> c6: left=2b23dcc6, right=9b7de2ee
--->> c7: left=9b7de2ee, right=84cdad86
--->> c8: left=84cdad86, right=2dbcda8d